\documentclass[fleqn,usenatbib]{mnras}

\usepackage{newtxtext,newtxmath}

\usepackage[T1]{fontenc}

\DeclareRobustCommand{\VAN}[3]{#2}
\let\VANthebibliography\thebibliography
\def\thebibliography{\DeclareRobustCommand{\VAN}[3]{##3}\VANthebibliography}

\usepackage{graphicx}	
\usepackage{amsmath}	
\usepackage{url} 
\usepackage{hyperref}

\title[Central galaxies alignment]{Alignment of the central galaxies with the environment}

\author[ Rodriguez, F., Merchán M. \& Artale, M. C.]{
Facundo Rodriguez,$^{1,2}$\thanks{E-mail: facundo.rodriguez@unc.edu.ar (FR)}
Manuel Merch\'an$^{1,2}$\thanks{E-mail: manuel.merchan@unc.edu.ar (MM)}
and M. Celeste Artale$^{3,4,5}$
\\
$^{1}$ CONICET. Instituto de Astronomía Teórica y Experimental (IATE). Laprida 854, Córdoba X5000BGR, Argentina.\\
$^{2}$ Universidad Nacional de Córdoba (UNC). Observatorio Astronómico de Córdoba (OAC). Laprida 854, Córdoba X5000BGR, Argentina.\\
$^{3}$ Physics and Astronomy Department Galileo Galilei, University of Padova, Vicolo dell’Osservatorio 3, I-35122, Padova, Italy\\
$^{4}$ INFN - Padova, Via Marzolo 8, I-35131 Padova, Italy\\
$^{5}$ Department of Physics and Astronomy, Purdue University, 525 Northwestern Avenue, West Lafayette, IN 47907, USA \\
}

\date{Accepted XXX. Received YYY; in original form ZZZ}

\pubyear{2022}

\begin{document}
\label{firstpage}
\pagerange{\pageref{firstpage}--\pageref{lastpage}}
\maketitle

\begin{abstract}
In this work, we combine ellipticity and major axis position angle measurements from the Sloan Digital Sky Server Data Release 16 (SDSS DR16) with the group finder algorithm of Rodriguez \& Merchán to determine the alignment of the central galaxies with the surrounding structures and satellite galaxies lying in their group. We use two independent methods: a modified version of the
two-point cross-correlation function
and the angle between the central galaxy orientation and the satellite galaxies relative position. The first method allows us to study the inner and outer regions of the cluster, while the second method provides information within the halos. 
Our results show that central galaxies present anysotropy in the correlation function up to $\sim 10 h^{-1}~{\rm Mpc}$,
which becomes $\sim$10\% stronger for the brightest ones ($^{0.1}M_{r}<-21.5$).
When we split the galaxy sample by colour, we find that red central galaxies are the main contributors to this anisotropy. We also show that this behaviour does not depend on the group mass or central galaxy ellipticity.
Finally, our results are in agreement with previous findings, showing that
the two-point cross-correlation function is a good tracer of the galaxy alignments using all galaxies and not only those of the group to which it belongs. In addition, this feature allows us to explore the behaviour of the alignment on larger scales.
\end{abstract}

\begin{keywords}
large-scale structure of Universe -- methods: statistical -- galaxies: haloes -- dark matter -- Galaxies: groups: general
\end{keywords}



\section{Introduction}

The current cosmological paradigm assumes that galaxies assemble by baryon condensation within the dark matter halos potential well \citep{White1978}.
In this context, to shed light on the dark matter nature, galaxy groups can be interpreted as a representation of dark matter halos to help our understanding of the halo-galaxy connection.

The way galaxies are distributed within dark matter haloes has imprinted crucial information on their large-scale assembly history and how environment impacts galaxy formation and evolution. At \textit{small-scales} observational reports show distinct features in the distribution of satellite galaxies which tend to be distributed within the galactic plane of the central galaxy \citep[see e.g.,][]{brainerd2005,Yang2005,Wang2008,Wang2010,Libeskind2015,Welker2018,Pawlowski2018}, and co-rotating it \citep{Ibata2001}. The preference for specific orientations between the central and satellite galaxies is also discussed by different works. Among them, \cite{Carter1980} and \cite{Binggeli1982} explore the alignment between the brightest cluster galaxies and the cluster they inhabit, and \cite{Struble1990} focus on the orientation of satellites galaxies and the brightest cluster galaxies. 

Other works show that satellite galaxies are preferentially distributed along the major axis of the central galaxy \citep{yang2006alignment, Azzaro2007,Wang2008}, and tend to be oriented towards the central galaxy of the group \citep{Pereira2005, Agustsson2006}. Also, the galaxy distribution shows alignment with the shape of the galaxy groups, and in consequence, with the shape of the dark matter haloes \citep{Paz2006,Lau2011,paz2011alignments, Smargon2012}.

Diverse authors show that red galaxies are better tracers of the orientation of their host halo, associated with their long mass accretion histories \citep[see e.g.,][]{Sales2004,yang2006alignment,Kang2007,Faltenbacher2007,Agustsson2010,Sarkar2022}. Among these contributions, we will follow \cite{yang2006alignment} that proposes the quantification of the alignment from the orientation of satellite galaxies relative to that of the major axis of the central galaxy of their group and computed the distribution of the angle concerning it. Using the redMaPPer cluster catalogue, \citet{Huang2016,Huang2018} find that the central galaxy cluster alignment is stronger for the redder, high-luminosity and size central galaxies. They also find that the alignment is stronger for the more elongated clusters. Their results show that satellites with redder colour, higher luminosity  tend to reside along the central galaxy major axis. 
Also \citet{Georgiou2019} confirm these trends using GAMA+KiDS survey.

From a theoretical point of view, within dark matter halos, internal tidal forces have induced galaxy alignments \citep{Ciotti1994,Usami1997,Porciani2002a,Fleck2003}. Instead, it could be primordial effects and large-scale tidal forces that handle the correlations between the shapes and spins of galaxies in different halos \citep{Pen2000,Catelan2001, Crittenden2001,Porciani2002b,Jing2002} jointly with the accretion of matter in preferential directions causing alignments with the filamentary structure of which dark matter halos are a part \citep{Faltenbacher2005,Bailin2005}. In agreement with the latter, several works in simulations relate the orientations of dark matter halos to the surrounding structures \citep[see e.g.,][]{Bond1996, Colberg2005, Kasun2005,Altay2006, Allgood2006, Basilakos2006,Aragon2007,Brunino2007,Bett2007,Hahn2007,Cuesta2008,Zhang2009, Noh2011}.

Also recently, hydrodynamical cosmological simulations have shown to play a fundamental role for understanding the intrinsic alignments between central galaxies, their satellites and the connection with the dark matter halo shape and orientation \citep[see e.g.,][]{Deason2011,Sales2012,Dong2014,Shao2016,Welker2018,Zjupa2020,Samuroff2021,Shi2021,Shi2021b,Tenneti2021}. These reports confirm that satellites are preferentially aligned with the major axis of the central galaxy, showing a stronger alignment between red satellites and red central galaxies. Furthermore, central galaxies tend to be aligned with the major axis of the dark matter halo hosts, with a stronger alignment for bright and spheroids. 
\citet{RagoneFigueroa2020} investigate the cosmic evolution of a subset of massive galaxy clusters, finding alignment between the clusters and the surrounding matter at large scales is present even at high redshifts, around $z\sim4$. They also find alignment between the brightest galaxy and the other galaxies in the cluster. These results suggest that galaxies tend to align in an outside-in fashion way, as a consequence of mass accretion in preferential directions \citep[see also,][]{Wang2009,Pandey2017}.

The fact that red satellite galaxies are strongly aligned with central red galaxies is described as \textit{anisotropic quenching} or \textit{angular conformity} \citep{Wang2008,Martin-Navarro2021,Stott2022}, and would be related with the well known \textit{galactic conformity} \citep[see e.g.,][]{Weinmann2006,Bray2016,Otter2020,Maier2022}. This phenomenon refers to the fraction of quiescent satellite galaxies around quenched central galaxies being higher than around star-forming central galaxies. Furthermore, different works also claim the existence of galactic conformity beyond the one-halo term \citep[see e.g.,][]{Kauffmann2013,Kauffmann2015,Lacerna2021}. The preferential alignemnts and abundance between quenched galaxies could be the manifestation of the large-scale structure formation and evolution of groups and clusters under different conditions, together with the modulation of different feedback processes within the haloes. 

A novel approach to quantify alignments in large-scale relevant to this work was proposed by \cite{paz2008angular}. This perspective is an adaptation of the traditional two-point cross-correlation function by comparing the clustering of different sub-regions with respect to a given orientation axis. This method has shown excellent performance in characterising the alignment of clusters/halos with surrounding structures in both simulations and observations \citep{paz2008angular,Faltenbacher2009,paz2011alignments, Faltenbacher2012, Schneider2012, Lopez2019}.

The galaxy group finder algorithm presented in \cite{rodriguez20} allows identification of systems in large galaxy surveys. This algorithm combines the friends-of-friends
\citep[FOF;][]{huchra} and halo-based methods \citep{Yang2005}. It has shown a high performance in the study of scaling relations between central and satellite galaxies \citep{rodriguez21}, in the mass assignment of the dark matter halos of the groups when compared with those obtained from weak gravitational lensing \cite{Gonzalez2021}, and in the studies of the halo occupation distribution in different environments \citep{alfaro2022}. In this work, we propose to identify a new galaxy group sample using the data provided by Sloan Digital Sky Survey Data Release 16 \citep[SDSS DR16,][]{ahumada2020} to study the alignment following the techniques introduced by \cite{yang2006alignment} and \cite{paz2008angular} and to compare the results with each other and with other authors.

Besides the significant contribution to our knowledge of the galaxy formation and dark matter distribution, understanding the features of galaxy alignments at small and large scales is relevant for weak gravitational lensing measurements since it is one of the most relevant physical systematic \citep[see e.g.,][]{Kirk2012,Troxel2015,Krause2016,Gonzalez2021b}. In particular, this is relevant for future surveys such as \textit{Euclid} \citep{EUCLID}, \textit{Nancy Roman Telescope}  \citep[WFIRST,][]{WFIRST}, and the Legacy Survey of Space and Time \citep{LSST}, constraining the cosmological parameters.

The structure of the paper is as follows. In Section \ref{sample}, we describe the SDSS 16 galaxy data and the group identification we use to measure the alignment. Analyses of the alignments using the cross-correlation function are presented in Section \ref{cf}. In section \ref{ang}, we study the distribution of satellite galaxies relative to the major axis of the central galaxy. Section \ref{groups ali} presents a comparison between the results using the central galaxy alignment and those obtained with the group shape alignment. Finally, Section \ref{concl} is devoted to discussing the implications of these results and providing a summary of the paper.
Throughout this work we adopt the standard $\Lambda$CDM cosmology \citep{Planck2016}, with parameters $\Omega _m$ = 0.3089, $\Omega _b$ = 0.0486, $\Omega _\Lambda$ = 0.6911, $H_0$ = 100h km s$^{-1}$Mpc$^{-1}$ with $h$ = 0.6774, $\sigma _8$ = 0.8159 and $n_s$ = 0.9667.

\section{Galaxies and groups}
\label{sample}

\begin{figure}
	\includegraphics[width=\columnwidth]{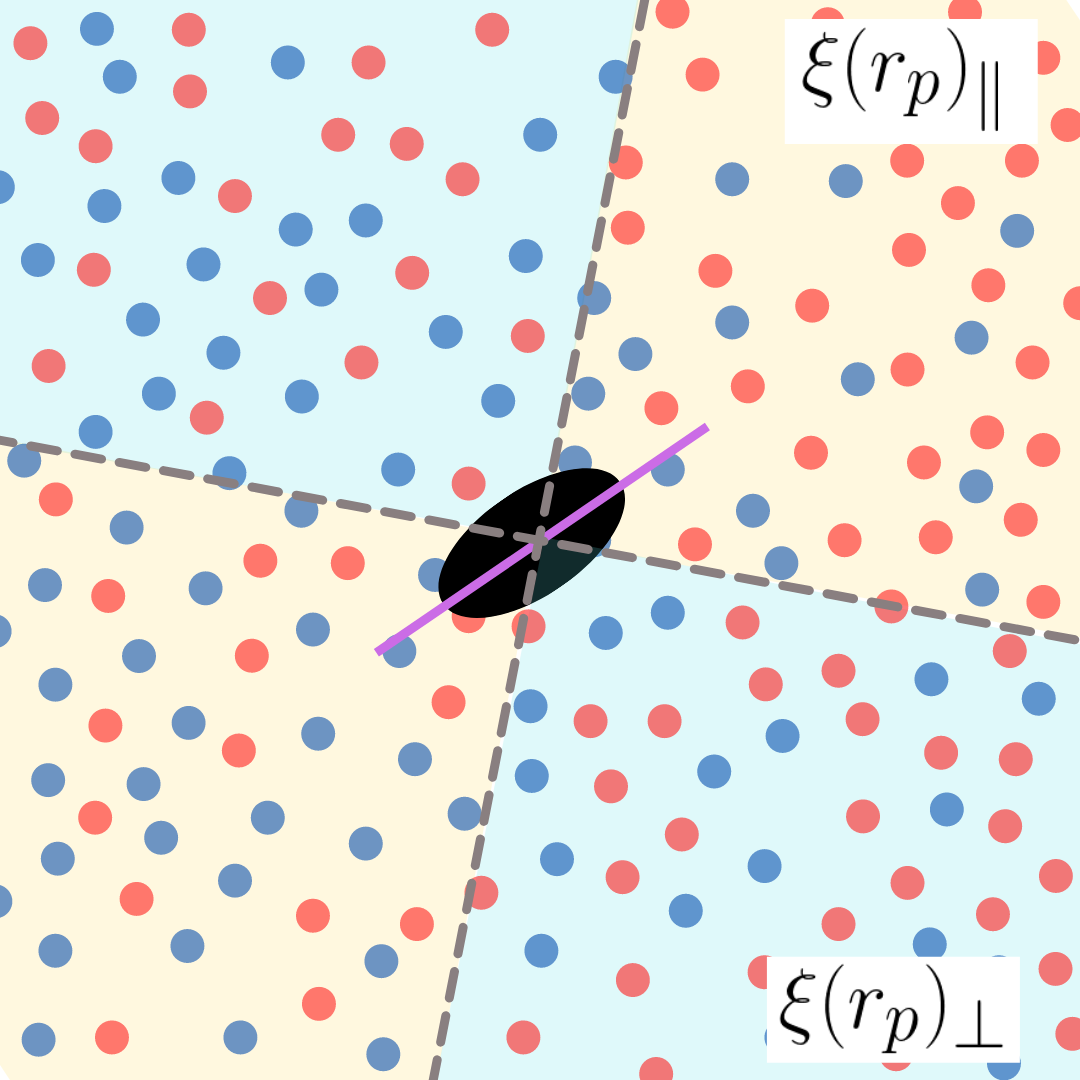}
    \caption{Schematic showing the regions used for the calculation of the anisotropic cross-correlation function parallel (yellow area) and perpendicular (light blue area) to the major axis of the central galaxy of the group (represented by the magenta line).}
    \label{esquema}
\end{figure}

\begin{figure}
	\includegraphics[width=\columnwidth]{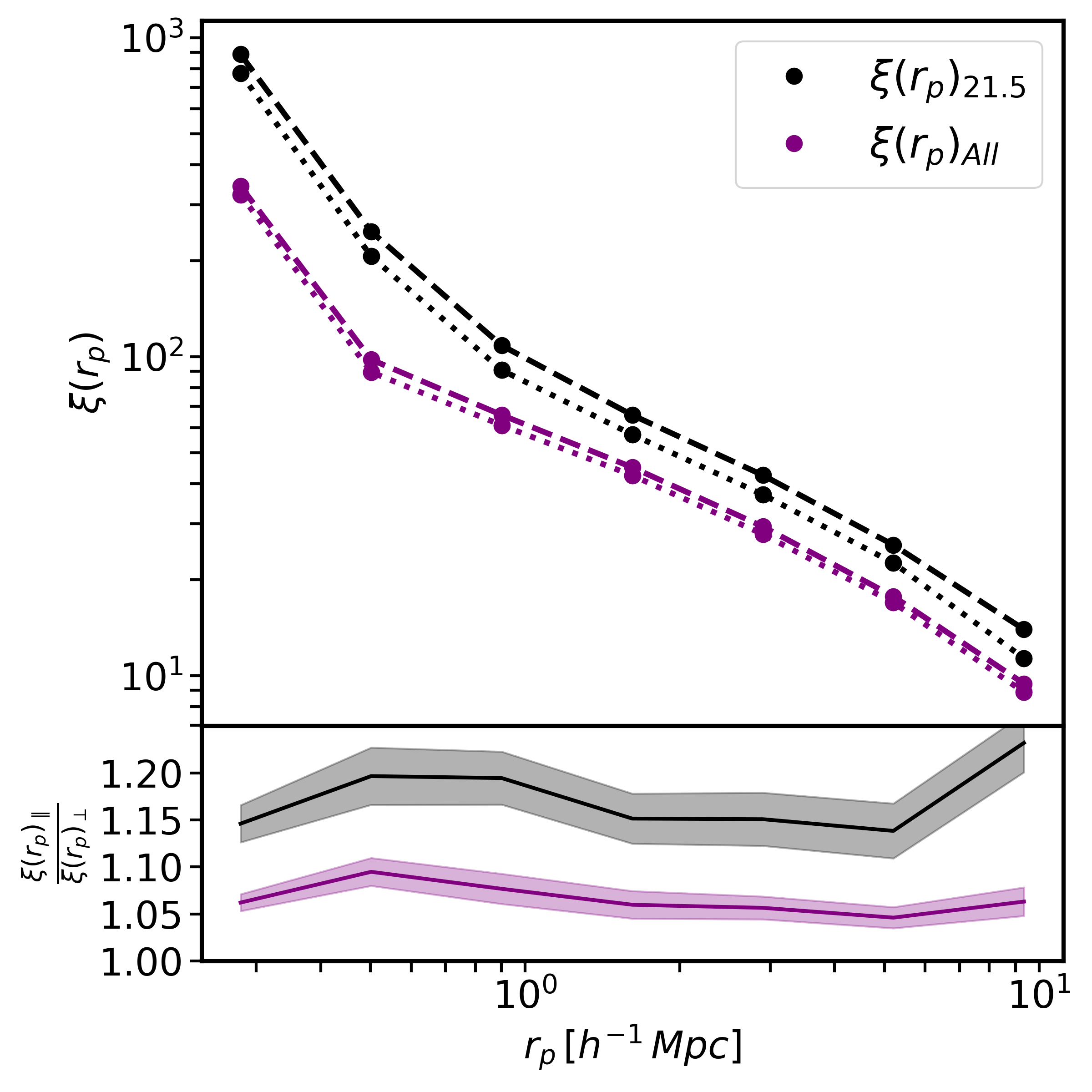}
    \caption{The anisotropic cross-correlation function using a total sample of central galaxies (violet) and those with $^{0.1}M_r<-21.5$ (black). The main panel shows the results for the parallel (dashed line) and perpendicular (dotted line) directions for both samples. The errors are computed with the jackknife method using a set of 50 subsamples and are typically smaller than the size of the points. The  subplot presents the fractional difference between the parallel and perpendicular direction. The shaded regions correspond to the error propagation from the correlation functions.}
    \label{fig:fc1}
\end{figure}

\subsection{Galaxy sample}
\label{sloan}

Throughout this work we will use a sample of galaxies obtained from the SDSS DR16 \citep{ahumada2020} both to extract the galaxy groups and to estimate the correlation function.
This survey is one of the largest publicly available photometric and spectroscopic galaxy catalogues. 
This is a cumulative survey that includes all the previous SDSS releases, containing observations covering roughly one third of the sky and includes more than two hundred million galaxies of which approximately 1.8 million have spectroscopic information. 
For our purposes, we keep those spectroscopic galaxies from the legacy footprint area with apparent magnitudes in r-band brighter than $17.77$ and redshifts lower than $0.2$.
Included in the large amount of information available for each galaxy, in addition to the positions and magnitudes, the most relevant for the analyses we will carry out are the ellipticity and the major axis position angle. 
These parameters are obtained from models based on the deVaucouleur or exponential profiles. As we find our results are independent of the model used, throughout this paper, we present only those achieve with the exponential model, although the figures would be very similar if we used the deVoucoleur model.

\subsection{Galaxy groups identification}
\label{data}

Following the procedure described in \cite{rodriguez20}, which combines both the FOF and halo-based methods, we extract galaxy groups from SDSS DR16. This method assumes that a group is a gravitationally bound system containing at least one bright galaxy. Thus, the procedure starts by identifying the groups using a percolation technique similar to the one described by \cite{huchra} adapted to fit this definition. Then, properties are assigned based on an estimate of the total luminosity of the group. In a second step, a halo-based iterative method comparable to the one developed by \cite{yang} is incorporated to improve the reliability of the system identification.  It is an iterative process, so that if the group membership changes, the properties of the dark matter halo are recalculated according to the new group luminosity. The process ends when there is no change in group membership. This method provides reliable galaxy systems with both low and high numbers of members. A detailed description of this algorithm and its performance can be found in \cite{rodriguez20}.

In addition to the basic group properties provided by the identification procedure (galaxy membership, mass, positions, etc), for the goal of this paper, we will need to classify central and satellite galaxies. To this end, we define the most luminous galaxy in each group as central and the remaining as satellites \citep{rodriguez21}. Given the particularities of this work, we select those groups with at least two galaxies.

\section{Correlation Function}
\label{cf}

\begin{figure*}
    \includegraphics[width=2\columnwidth]{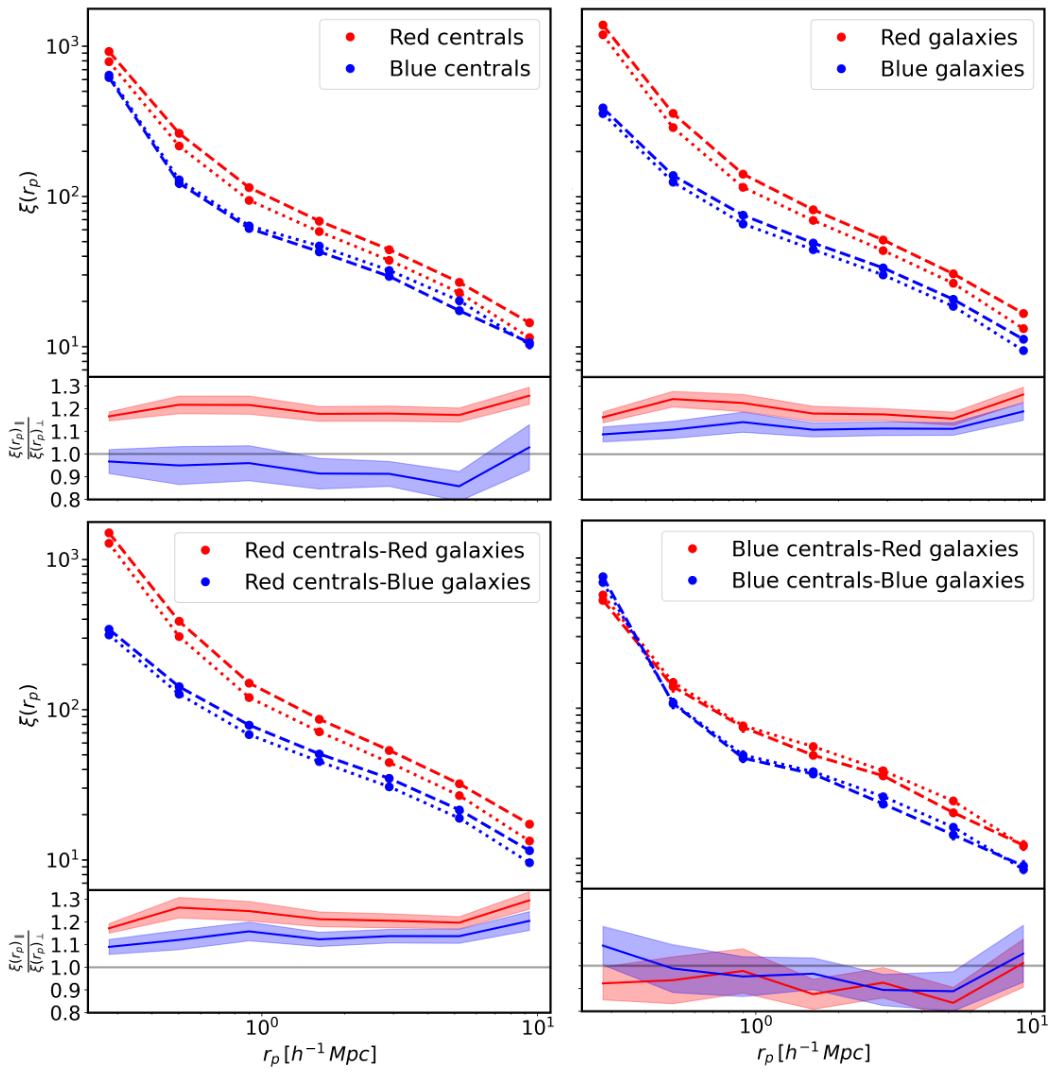}
    \caption{
    The anisotropic cross-correlation function in the same format as Figure \ref{fig:fc1} for the central galaxies sample with $^{0.1}M_r<-21.5$ split by colour. The top left panel shows our results for the red and blue central galaxies, while the top right panel shows those for target red and blue galaxies. We also study the anisotropic cross-correlation function of the red centrals with the target population split by color (bottom left panel). The same analysis is done for the blue centrals on the bottom right panel.
    }
    \label{fig:fc2}
\end{figure*}

\begin{figure}
	\includegraphics[width=\columnwidth]{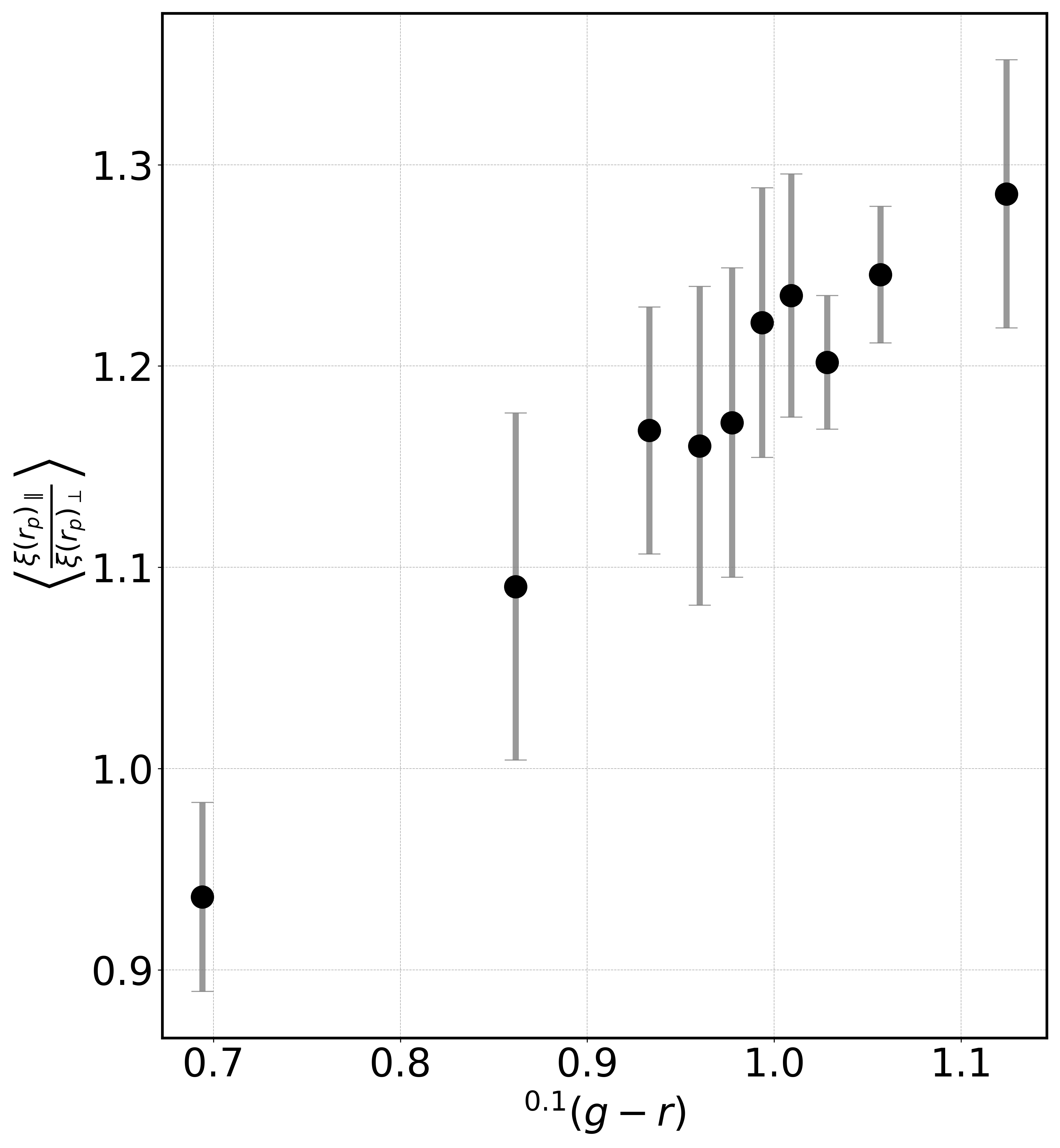}
    \caption{ Mean anisotropy as a function of the colour of the central galaxies, divided into deciles. 
    The distribution of the points in the $^{0.1}(g-r)$ axis is irregular due to the colours of the central galaxy sample being concentrated around $\sim 1$. Error bars correspond to the standard deviation of the  ${\xi(r_p)_{\parallel}}/{\xi(r_p)_{\perp}}$ values between 0.1 and 12 $h^{-1}Mpc$.}
    \label{fig:colorani}
\end{figure}

\begin{figure*}
	\includegraphics[width=2\columnwidth]{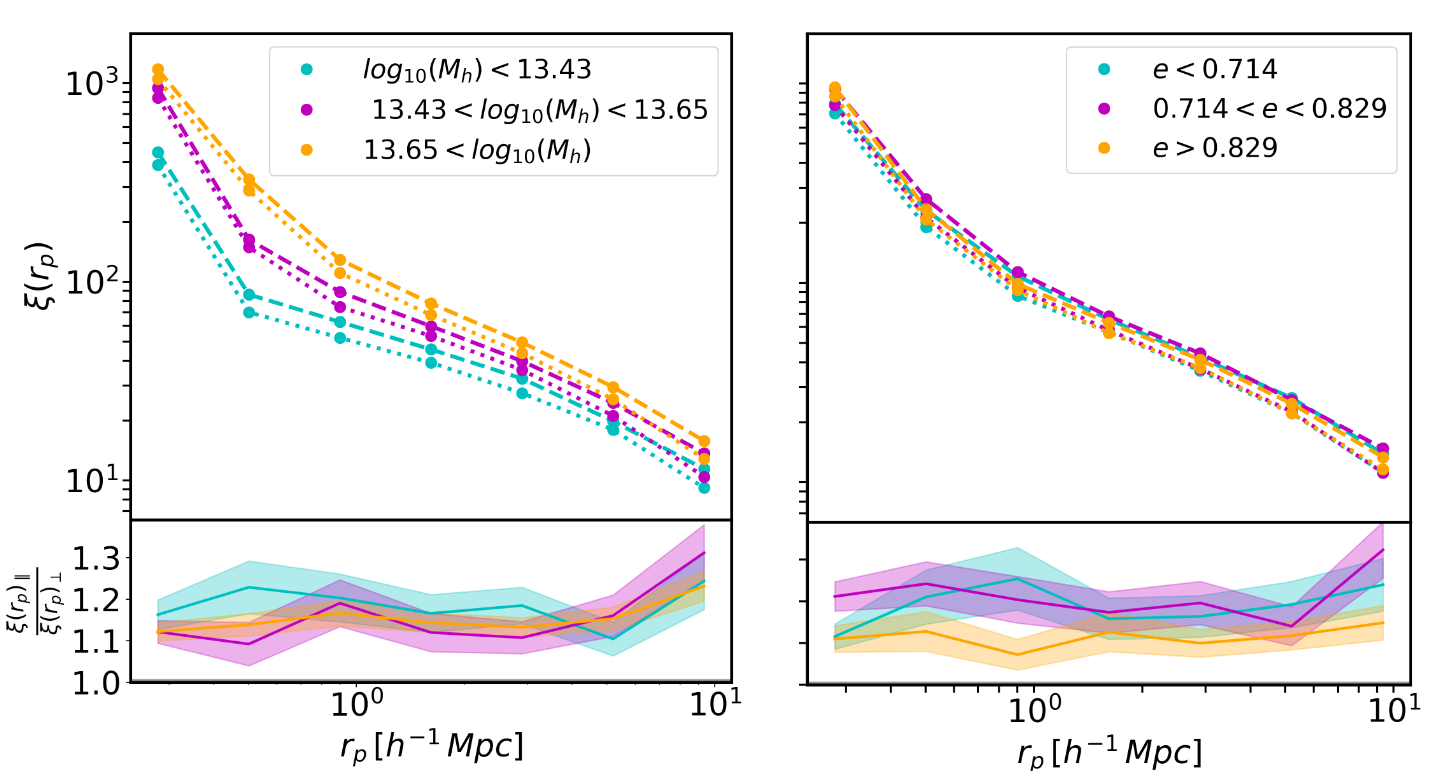}
    \caption{The anisotropic cross-correlation function in the same format as Figure \ref{fig:fc1} for the sample of central galaxies with $^{0.1}M_r<-21.5$ split by group mass (left panel) and ellipticity (right panel) terciles.}
    \label{fig:fc3}
\end{figure*}

\begin{figure}
	\includegraphics[width=\columnwidth]{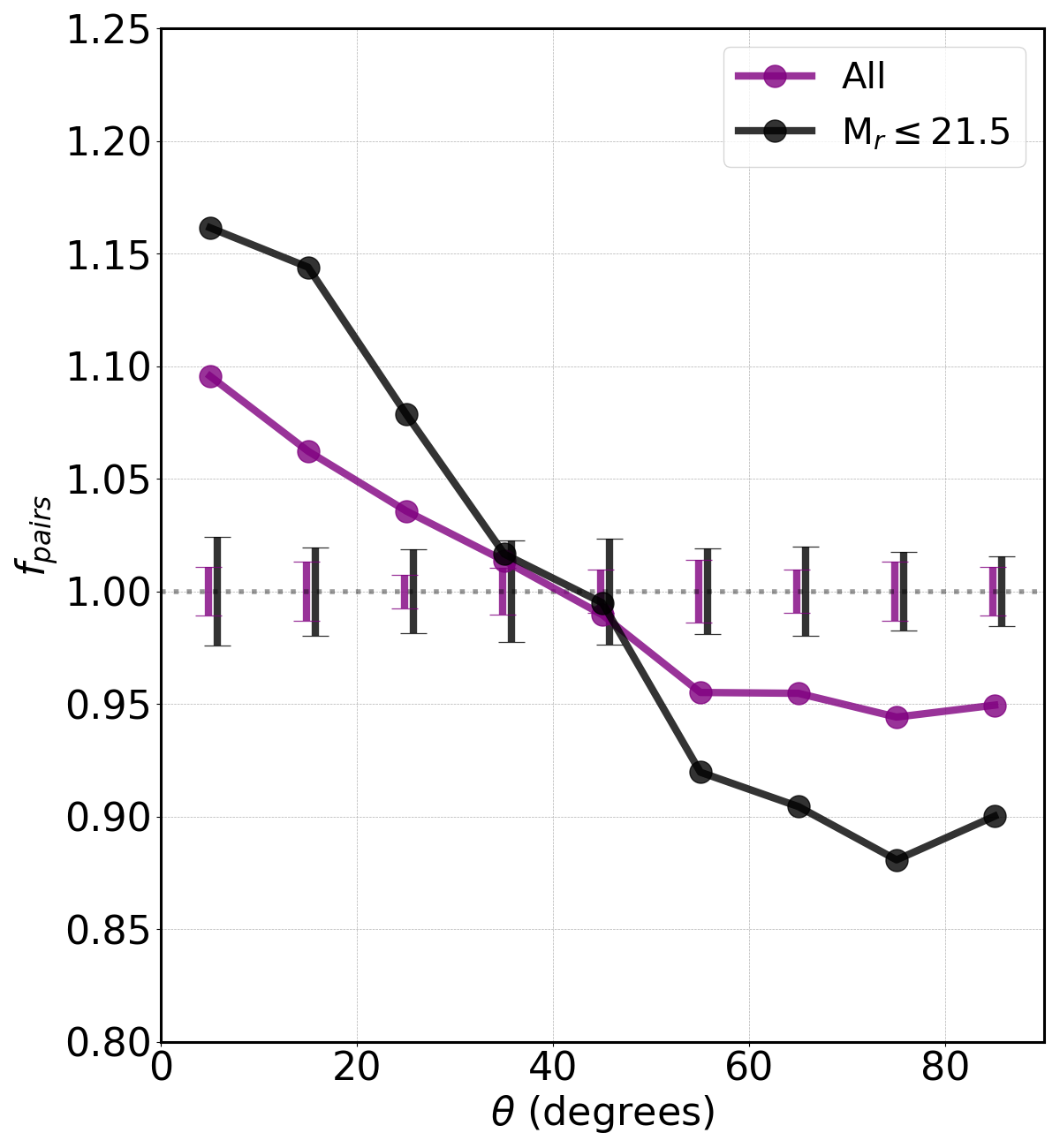}
    \caption{The normalized probability distribution of the angle between the orientation of the major axis of the central group galaxy and the direction of each satellite as measured from the central galaxy, using the total sample of central galaxies (violet) and those with $^{0.1}M_r<-21.5$ (black). We performed the calculations following the technique presented by \citet{yang2006alignment}. 
    We compute the significance of the deviation (error bars) by using 100 random samples for the orientation of central galaxies.
    }
    \label{fig:ang1}
\end{figure}

To statistically study the distribution of galaxies in different directions, we will use a modified version of the cross-correlation function between central galaxies of groups and the total galaxy sample  (target galaxies).
This function measures the excess probability with
respect to a random distribution, that a galaxy  will reside at a distance $r$ away from a given central galaxy.
To avoid the redshift space distortions effects,
we estimate the correlation function in two coodinates,
one parallel ($\sigma$) and one perpendicular ($\pi$) to the line of sight ,$\Xi (\sigma, \pi)$, and integrating over the $\pi$ direction. 
The result from this integration is the projected
correlation function: 
\begin{equation}
    \xi(r_p)=2 \int_{\pi_{min}}^{\pi_{max}} \Xi(\sigma,\pi) d\pi
    \label{eq1}
\end{equation}
where $\pi_{min}=0.1h^{-1} Mpc$ and $\pi_{max}=12h^{-1} Mpc$. 
To estimate $\Xi (\sigma, \pi)$ we use the standard estimator,

\begin{equation}
\Xi(\sigma,\pi) = \frac{N_{ran}}{N_{gal}} 
\frac{N_{CG}(\sigma,\pi)}{N_{CR}(\sigma,\pi)}-1
\end{equation}
where  CG and CR denote Central galaxy-Galaxy and Central galaxy-Random pairs, repectively. $N_{gal}$ is the number of galaxies and $N_{ran}$ is the number of points in a random catalogue. These random catalogues were built assigning random angular positions with a uniform distribution (inside the SDSS DR16 area) and mimicking the redshift distribution of each sample. We made samples random with different object numbers and, to ensure the reduction of uncertainties introduced by the Poisson random noise, throughout this work, we chose $N_{ran}\sim 50 N_{gal}$. These quantities allow us to keep the samples large enough to reduce the noise and to have the computation times reasonable.

\subsection{Anisotropic correlation function}

As mentioned above, the main objective of this work is to determine whether galaxies tend to align with respect to the central galaxies of the groups.
To study the alignment of the galaxy distribution with respect to a particular direction, we follow the same approach as \cite{paz2008angular}. 
For this, the cross-correlation is computed by counting pairs depending on the angle between the galaxy position relative to the central galaxy and its major axis (provided by SDSS DR16, as referred in section \ref{sloan}).
We define $\xi(r_p)_{\parallel}$ as the correlation
function calculated taking into account all pairs 
within an angle of $45^{\circ}$ (light yellow region, 
Figure~\ref{esquema}) while the correlation function calculated with the pairs with an angle greater than 
$45^{\circ}$  (light blue region, Figure~\ref{esquema}) is defined as $\xi(r_p)_{\perp}$.

Using the groups described in section \ref{data} we made 
a first measurement of $\xi(r_p)_{\parallel}$ and
$\xi(r_p)_{\perp}$. For this purpose, we chose those groups with at least two member galaxies and with redshift $z<0.2$. This result is shown in purple in Figure \ref{fig:fc1}. 
In this figure and the rest of our analysis, we compute the errors with the jackknife technique. However,  in many cases, the error bars are smaller than the points used for the correlation function.
We decided to restrict the sample of groups to those whose central galaxy is brighter than $^{0.1}M_r<-21.5$ (remaining $\sim 14000$ groups) and, naturally, as the central galaxy gets brighter, the mass of the group increases and, therefore, the anisotropy is expected to be larger \citep{paz2011alignments}. The result for this sample is also shown in Figure \ref{fig:fc1} (black symbols). As can be seen, the correlation increases at the same time as the anisotropy which reaches values up to 15\%. This will allow us to analyse the dependence of the anisotropy with different properties.

It is worth noting that the anisotropy signal manifests itself with similar intensity in both the one-halo and two-halo terms. As is known, dark matter halos show an alignment of the major axis with the large-scale structure \citep[e.g.,][]{paz2011alignments}. The large scale anisotropy signal could therefore be associated with an alignment of the central galaxy with the halo shape.
We will deal with this feature later.
The effect of the projection in the equation \ref{eq1} must also be taken into account because there are contributions from galaxies that are actually at different distances, particularly on small scales.

\subsection{Colour dependence}
\label{fccolordependence}

First, we will analyse the dependence of the anisotropy of the correlation function on colour. For this purpose, we define red and blue galaxy samples using as colour indicator $^{0.1}(g-r)$, which corresponds to the $k$-corrected colour to redshift $z = 0.1$ using Sloan photometric bands and $k$-correction calculated using the 4\_3 version of the \emph{k-correct} software described in \cite{blanton2007k} and kindly made available by the authors.  We will define as red (blue) galaxies those with  $^{0.1}(g-r)>0.83$. This threshold value is associated with the bimodality of the colour-magnitude relation \citep{yang2006alignment,weinmann2006properties}.

Following the same criteria as \cite{yang2006alignment}, we explore the dependence of anisotropy on colour, both for the central and target galaxies. Using the absolute magnitude cut described above ($^{0.1}M_r<-21.5$), we want to assess which galaxies are responsible for the anisotropy observed in Figure \ref{fig:fc1}, i.e. whether it is the colour of the central galaxy or the environment that determines the overall anisotropy signal.

Following the criteria described above, we made samples by selecting red and blue galaxies for both centres and targets. All these samples were combined to analyse the origin of the observed anisotropy signal. In Figure~\ref{fig:fc2}, we show in an organised manner the correlation functions corresponding to the combination of each of these samples. It should be noted that the fraction of blue central galaxies is significantly smaller than the red ones (13195 red and 1705 blue), which is not the case for the total sample of galaxies, where  $\sim$ 40\% are blue. That is why errors are larger for central blue samples. Moreover, as expected, the correlation functions associated with the red galaxy samples are always higher than the blue ones .

Since the aim of this section is to find out which type of galaxies are responsible for the anisotropy, the first issue that can be deduced from the Figure \ref{fig:fc2} is that the anisotropy signal is dominated by those groups whose central galaxies are red. When only the colour of target galaxies is taken into account (top right panel), it can be seen that both blue and red galaxies show a significant and nearly indistinguishable anisotropy signal. However, when the colour differentiation is performed on the central galaxies (top left panel), the anisotropy signal corresponding to blue galaxies disappears while  red galaxies maintains an amplitude $\sim$20\% above the blue ones, with a statistical significance greater than two sigma over the whole range. 
The difference in the correlation, in this case, is because the groups with red central galaxies are more massive than those with blue central galaxies.

To further highlight the influence of central red galaxies on the anisotropy signal, we cross-checked the central and target galaxy samples. Again, what is observed is that for the central red galaxies the anisotropy feature is still present regardless of the colour of the target galaxies (lower left panel). In contrast, for blue central galaxies, the anisotropy signal vanishes.

So far, we have shown that the colour of the central galaxies is decisive for the anisotropy signal. 
However, by imposing a binary selection, the number of red central galaxies is greater than the number of blue ones. Thus, to analyze in detail the impact that colour has on the anisotropy signal, we divide the sample of central galaxies into colour deciles containing $\sim 1400$ galaxies.
For each subsample, we compute the parallel and perpendicular correlation functions and the corresponding anisotropy in the same way as in the previous analyses. To show the colour dependence clearly, we calculate for each decile the average anisotropy signal over the whole range of distances ($\left<{\xi(r_p)_{\parallel}}/{\xi(r_p)_{\perp}}\right>$). As shown in Figure \ref{fig:colorani}, we find that the anisotropy signal increases gradually as the central galaxies become redder. 

As in the general case, anisotropy manifests in the same way at all scales regardless of colour. As already mentioned in the instance of the overall correlation function, different physical processes are involved at different scales. Therefore, one would expect in the colour analyses a different behaviour between the regimes corresponding to the 1 and 2 halo terms. However, this is not observed. To study this behaviour, we will further explore the alignment of the central galaxy with the shape of the group according to its colour.

\subsection{Mass and ellipticity dependence}
\label{fcmedependence}

We now proceed to study the dependence of the anisotropy on the ellipticity of central galaxies provided by the SDSS DR16 and the group masses determined in the identification process.

Given that the more massive the groups are, the higher the correlation, it is natural to ask whether there is a dependence of the anisotropy of the correlation function on the mass when the shape of the central galaxy is taken as a reference. This analysis is further justified if one takes into account that, from numerical simulations, \cite{paz2011alignments} found that there is a strong dependence of the anisotropy of the correlation function when using the shape of the haloes as a reference. However, the same authors do not detect this trend in observations.

The left panel of Figure~\ref{fig:fc3} shows the correlation function and its anisotropy for three mass ranges divided into terciles. As can be seen, while the height of the correlation function increases with mass as expected, the anisotropy does not show any dependence on the group masses considering the estimated errors.
The right panel of Figure~\ref{fig:fc3} shows the ellipticity of the central galaxy sample split into terciles. Taking into account the estimated errors, no dependence on anisotropy is observed but, unlike the mass,  the strength of the correlation function is similar for all samples. 
The similitude in the anisotropy signal in both the mass and ellipticity samples is because in all cases the percentage of red central galaxies is greater than 80\% and similar to that of the total sample (Figure~\ref{fig:fc1}).

\cite{paz2011alignments} found, from numerical simulations, a strong dependence of the anisotropy on the correlation function with mass when using the shape of the haloes as a reference. 
We take as a reference the shape of the central galaxy. If they were aligned with the halo shape in which they reside, the anisotropy signal should depend on group mass. Our results do not reflect this dependence. However, the same authors do not detect this trend in the observations, showing that projection effects are important. Moreover, the hypothesis that the central galaxies are strongly aligned with the haloes may not be plausible.

\section{Central-satellite angular distribution}
\label{ang}

\begin{figure*}
	\includegraphics[width=2\columnwidth]{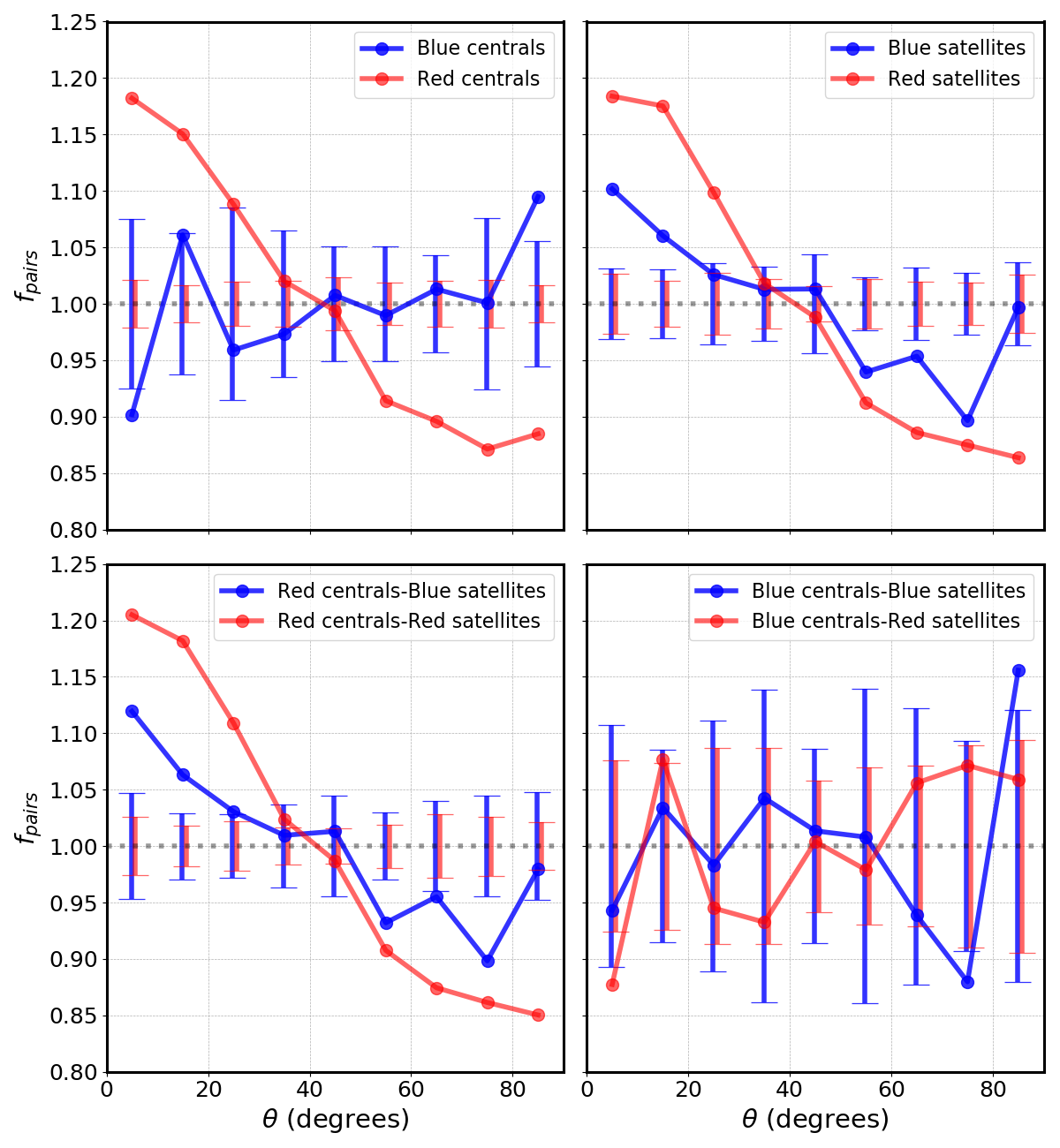}
    \caption{
    The normalized probability distribution of the angle between the major axis of the central galaxy and the direction of each satellite in the group measured from the central galaxy, $f_{pairs}$. In all panels, we use the groups whose central galaxies have $^{0,1}M_r<-21,5$. 
    We split the galaxy sample into red and blue central galaxies (top left panel), and into red and blue satellites (top right panel).
    The bottom left panel shows our results for the red-central galaxies split by red and blue satellite galaxies, and the bottom right shows the blue-central ones divided by red and blue satellite galaxies.}
    
    \label{fig:ang2}
\end{figure*}

\begin{figure*}
    \includegraphics[width=2\columnwidth]{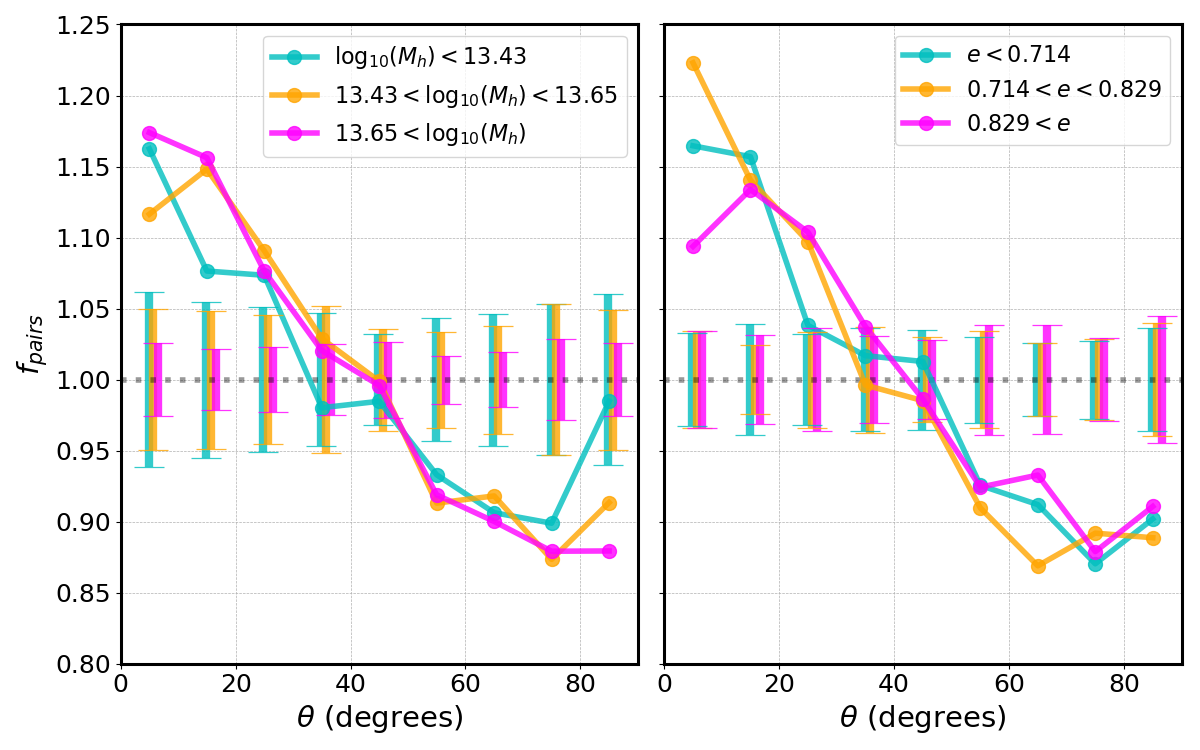}
    \caption{
    The normalized probability distribution of the angle between the major axis of the central galaxy and the direction of each satellite in the group measured from the central galaxy using groups whose central galaxies have $^{0.1}M_r<-21.5$ (black) and separating the
    samples into group mass (left panel) and  ellipticity (right panel) terciles.}
    \label{fig:ang3}
\end{figure*}

An alternative approach to the study of the alignments between the central galaxy and its environment is to statistically analyse the angle that the central galaxy position subtends the direction of the other galaxies in its group. More specifically, we will use the method developed by \cite{yang2006alignment}, where only galaxies of the same group are considered. This will not only give us an independent estimate but also allow us to compare the above results with those of \cite{yang2006alignment} .

According to these authors, the alignment between the major axis of the central galaxy and the relative direction of the other galaxies in the group can be quantified by normalised pair counting:
\begin{equation}
    f_{pairs}(\theta)=\frac{N(\theta)}{\left< N_{r}(\theta)\right>}
\end{equation}
where  $N(\theta)$ is the total number of central-satellite pairs for a given value of $\theta$, and $\left< N_{r}(\theta)\right>$ is the mean number of pairs obtained from 100 samples with the randomised orientations of the central galaxies.
With this definition $f_{pairs}(\theta)=1$ for a random sample while the deviation from unity quantifies the intensity of the alignment. To determine the significance of this deviation, we compute $\sigma_r(\theta)/\left< N_{r}(\theta)\right>$, where $\sigma_r(\theta)$ is the standard deviation of $N_{r}(\theta)$ obtained from the 100 random samples.

To compare with previous results, we use the same selection criteria as in the previous section but applying the method proposed by \cite{yang2006alignment}. It should be noted that, unlike the correlation function, only member galaxies are used for this statistic.
So the first result corresponds to the comparison between the group sample without considering the brightness of the central galaxy and the one whose central galaxies are brighter than $^{0.1}M_r<-21.5$. These results are shown in Figure \ref{fig:ang1}, where it can be seen that, in the same way as in Figure \ref{fig:fc1}, the alignment increases when the central galaxies are brighter. Moreover, the values obtained with this sample are in excellent agreement with the \cite{yang2006alignment} results.

The technique described above was applied to the colour-selected samples presented in section \ref{fccolordependence}. Figure~\ref{fig:ang2} shows the results obtained for the colour cross-samples between central and satellite. 
In general, as obtained employing the correlation function, it is observed that the alignment is produced mainly by the red central galaxies.  Note that when we do not differentiate the central galaxies by colour (upper right panel), a significant signal is observed for both red and blue satellite galaxies, although the blue ones have a smaller signal. However, when we distinguish the central galaxies by colour, the signal is only observed for the red sample (top left panel). This effect is confirmed in the cross-samples (bottom panels), in which the signal only appears for the red central galaxy sample while the results corresponding to the blue central galaxies lies within the error bars. 

It is also important to highlight that these results are consistent with both the previous results and those of \cite{yang2006alignment}. However, this type of analysis is restricted to galaxies residing in the same halo, whereas the correlation function involves large-scale structure, allowing us to study the anisotropy on larger scales.  

In the instance of the samples differentiated by mass and ellipticity (left and right panels of the figure \ref{fig:ang3}, respectively), as observed in section \ref{fcmedependence}, there is no alignment dependence on these properties. Although the mass and ellipticity intervals are slightly different from those of Yang, the results are consistent.

\section{Central and group alignment}
\label{groups ali}

\begin{figure}
	\includegraphics[width=\columnwidth]{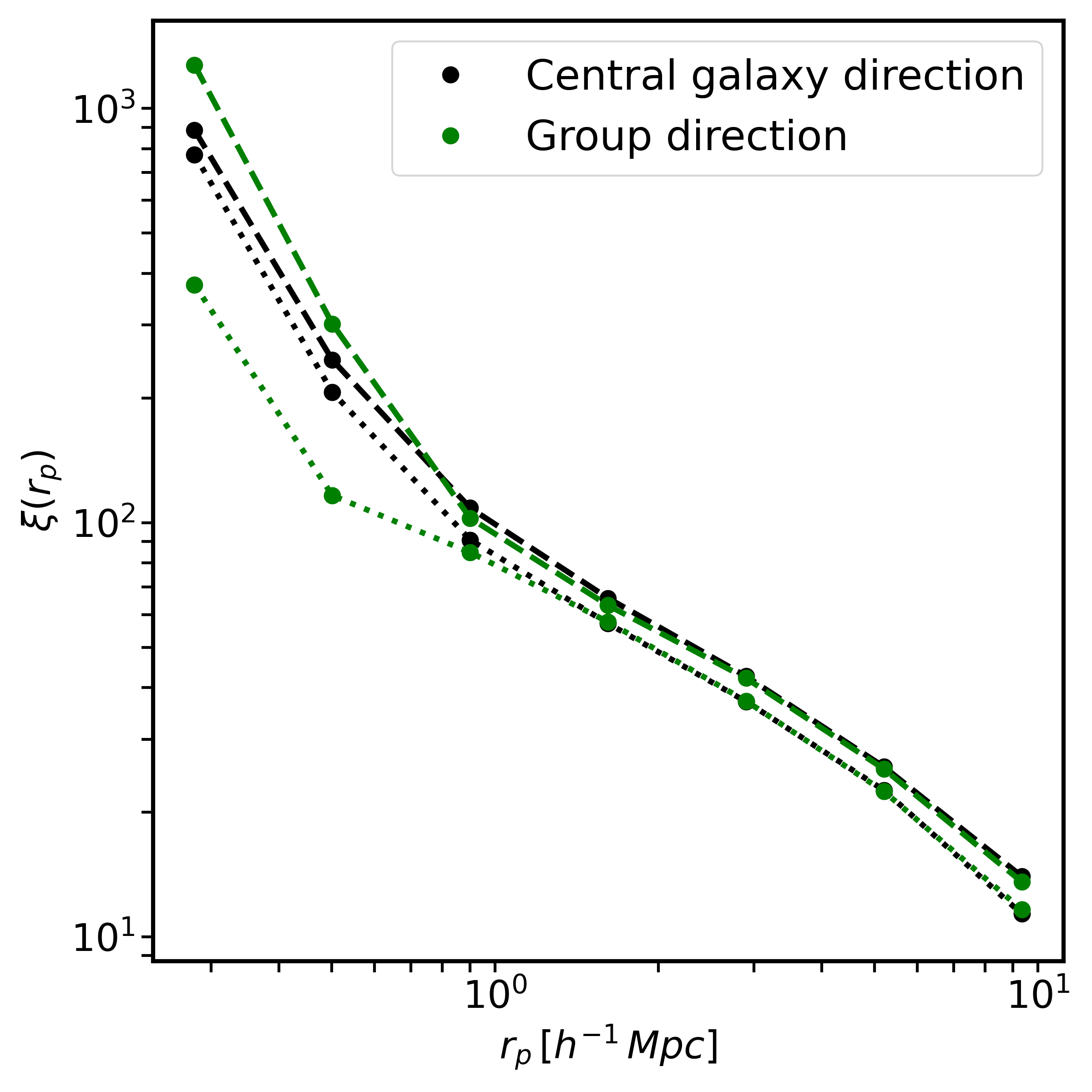}
    \caption{The comparison between the anisotropic cross-correlation function using the sample of central galaxies with $^{0.1}M_r<-21.5$ in the direction parallel and perpendicular to the main axis of the central galaxy presented in figure \ref{fig:fc1} (black) and the main axis of the shape of the galaxy group in which they reside (green).}
    \label{fig:fc4}
\end{figure}

\begin{figure}
	\includegraphics[width=\columnwidth]{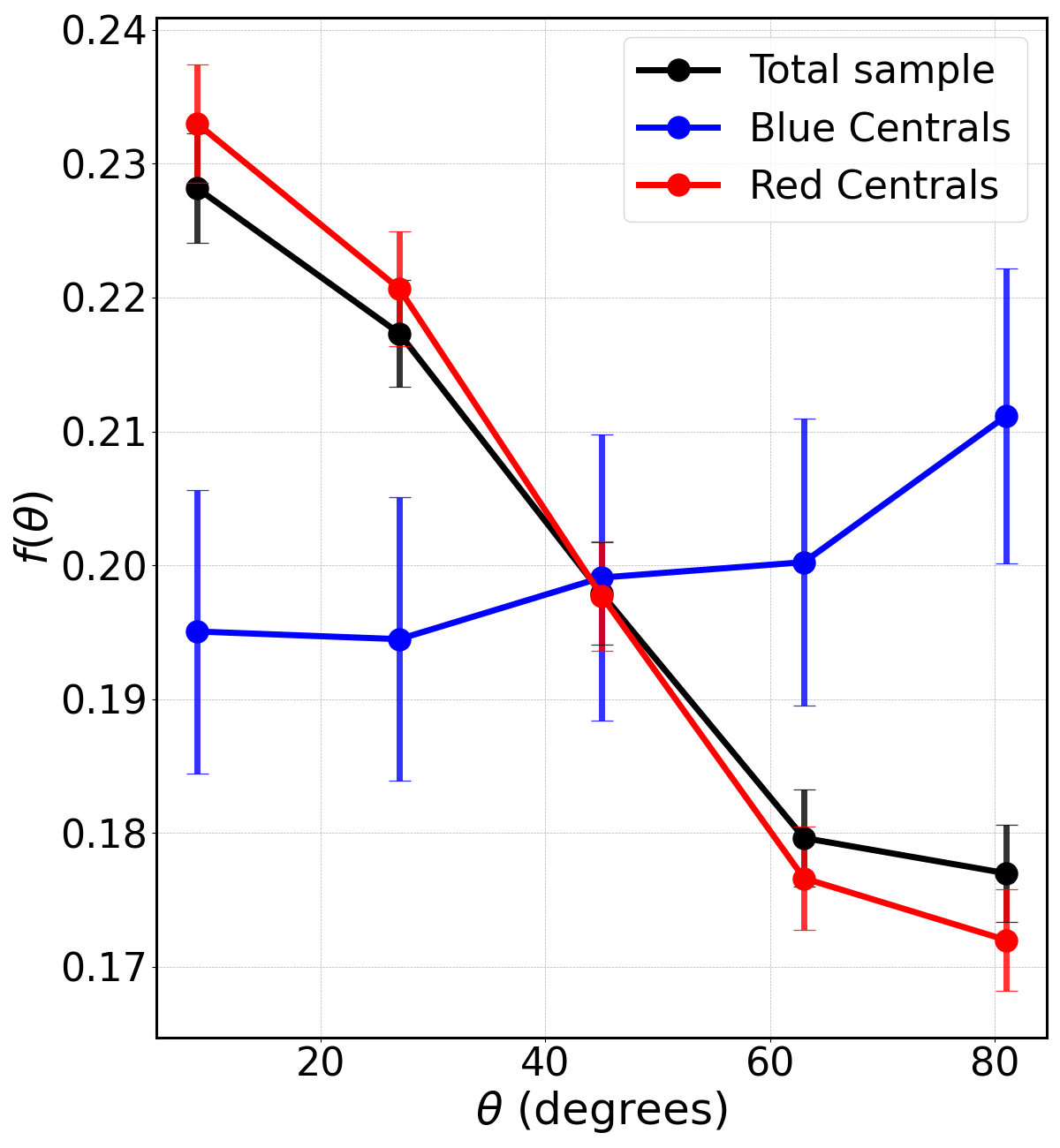}
    \caption{The measured probability density distribution of the alignment between the central galaxy and the group shape for the total sample of central galaxies with $^{0.1}M_r<-21.5$, and separating these central galaxies by colour.}
    \label{fig:ang4}
\end{figure}

As seen in section \ref{cf}, the cross-correlation function between central and target galaxies shows a significant anisotropy on large scales (mainly for red central galaxies). This effect is hard to explain since, on these scales, the physical processes involved are unlikely related to those linked to the evolution of the central galaxy. On the other hand, some work shows that the large-scale structure is aligned with the shape of dark matter halos and galaxy groups \citep{paz2011alignments}. This leads us to speculate that there is an alignment between the central galaxy and the group to which it belongs. 

As a first test, to evaluate our hypothesis, we estimate the anisotropic cross-correlation function but, in this case, aligned with the direction of the group shape major axis. This direction was calculated from the eigenvector associated with the largest eigenvalue of the shape tensor 
\begin{equation}
    I_{ij}= \frac{1}{N_g}\sum_{\alpha=1}^{N_g} X_{\alpha i} X_{\alpha j},
\end{equation}
where $X_{\alpha i}$ is the $i^{th}$ component of the Cartesian coordinates in the plane of the sky of a galaxy $\alpha$ relative to the centre of the group , and $N_{g}$ is the number of galaxies in the group.
Figure \ref{fig:fc4} shows the result obtained using the group shape and the corresponding to the central galaxies presented in  figure \ref{fig:fc1}. The one halo term for the groups shows a much larger anisotropy than the central galaxies, i.e. the central galaxies are not as aligned with the group as we thought. But, strikingly, the anisotropies in the two-halo term look very similar.

To quantify the alignment between the central galaxy and the group shape we measured the angle between the corresponding major axes. The figure \ref{fig:ang4} shows the measured probability density distribution for the total, red and blue central galaxy samples. As can be seen, the total sample shows a not very pronounced alignment in line with the results obtained for the one-halo term of the earlier analysis. Only $\sim 54\%$ have an alignment angle smaller than $45^{\circ}$. Furthermore, in agreement with all the results of the previous sections, the red central galaxies are responsible for the signal, while the blue central galaxies do not show any alignment.

\section{Conclusions}
\label{concl}

From a catalogue of galaxy groups extracted from the SDSS, we used the two-point cross-correlation function to analyse the anisotropy of their environment by taking the orientation of the central galaxy as a reference. In this way, we explored both the inner regions of galaxy groups and the larger scales associated with the surrounding structures. We complemented these studies with the technique proposed by \cite{yang2006alignment}, which allowed us to statistically analyse the angle between the orientation of the central galaxy of the group and the relative positions of the other member galaxies.

Despite the projection effects that distort the real shape of the galaxies, as a first significant result, we can conclude that there is a remarkable anisotropy of the correlation function when we take as a reference the orientation of the central galaxy, particularly if we consider those groups with bright central galaxies.

When we select samples by colour, the result that emerges is that the main contributors to the anisotropy are the red central galaxies, whereas the blue ones do not make any contribution to the signal. This effect may be because the red central galaxies reside in older environments, so the processes they have undergone throughout their history may have contributed to the observed difference in anisotropy between red and blue central galaxies. On the other hand, when selecting samples in mass or ellipticity terciles, we do not observe any correspondence of the anisotropy with these variables. However, other works show that anisotropy increases with mass which, in turn, is associated with the shape. Therefore, as the projection effects could be hiding some dependence, it would be interesting to deepen this analysis in numerical simulations.

When we applied \citeauthor{yang2006alignment}'s technique, we obtained results in agreement with these authors and consistent with those obtained using the correlation function. These results reinforce those observed on scales within the halo but do not add any information regarding anisotropy on larger scales given that only group members galaxies are used in this analysis. 

The anisotropies observed on large scales led us to presume an alignment between the central galaxy and the group it inhabits. We then calculated the correlation function taking as reference the shape of the group and the probability distribution function of the angle between the position of the central galaxy and the group orientation.  As expected, the one-halo term shows a large anisotropy (as it is well known, haloes tend to be prolated). Even though this is not a novel result,  it was obtained from our catalogue of groups, which brings a new validation to the implemented identification technique. The estimated observed anisotropy related to the direction of the major axis of the group is considerably larger than that associated with the central galaxy, suggesting that the group and its central galaxy are not closely aligned. Consequently, we study the probability distribution function of the angle between the major axes of the central galaxy and that of its group. We find a weak overall alignment dominated by the central red galaxies.

Finally, we would like to point out that our way of measuring anisotropy not only allows us to study the internal regions of the group but also to analyse larger scales. It is also important to note that the anisotropy associated with the two-halo term is similar with respect to both the shape of the group and the shape of the central galaxy. This result is difficult to explain considering that the central galaxy is poorly aligned with its group. It could be that there is another direction more than the ones studied in which the anisotropy is stronger. Furthermore, this work was carried out with observational data and, consequently, affected by projection effects. We, therefore, consider it would be worthwhile to reproduce this work on numerical hydrodynamic simulations.

\section*{Acknowledgements}
The authors wish to thank the anonymous referee for his/her report that helps us to improve this manuscript.
FR and MM thanks the support by Agencia Nacional de Promoci\'on Cient\'ifica y Tecno\'ologica, the Consejo Nacional de Investigaciones Cient\'{\i}ficas y T\'ecnicas (CONICET, Argentina) and the Secretar\'{\i}a de Ciencia y Tecnolog\'{\i}a de la Universidad Nacional de C\'ordoba (SeCyT-UNC, Argentina).
MCA acknowledges financial support from the Seal of Excellence @UNIPD 2020 program under the ACROGAL project.
\section*{Data Availability}
 The data underlying this article will be shared on reasonable request to the corresponding authors. 


\bibliographystyle{mnras}
\bibliography{main} 



\bsp	
\label{lastpage}
\end{document}